\documentclass[sigconf]{acmart}

\usepackage{booktabs} 
\usepackage{enumitem}
\usepackage{balance}
\usepackage[utf8]{inputenc}
\usepackage{multirow}
\usepackage{graphicx}
\usepackage{microtype}
\usepackage{subfigure}
\usepackage{bm}
\usepackage{amsmath}
\usepackage{amsfonts}
\usepackage{graphicx}

\AtBeginDocument{%
  \providecommand\BibTeX{{%
    \normalfont B\kern-0.5em{\scshape i\kern-0.25em b}\kern-0.8em\TeX}}}

\setcopyright{acmcopyright}
\setcopyright{rightsretained}
\copyrightyear{2018}
\acmYear{2018}
\acmDOI{10.1145/1122445.1122456}

\begin{document}
\title{Large Language Models as Data Augmenters for \\Cold-Start Item Recommendation}



\author{Jianling Wang, Haokai Lu, James Caverlee, Ed Chi and Minmin Chen}
\affiliation{%
  \institution{Google DeepMind}
}
\email{{jianlingw, haokai, caverlee, edchi, minminc}@google.com}






\begin{abstract}

The reasoning and generalization capabilities of LLMs can help us better understand user preferences and item characteristics, offering exciting prospects to enhance recommendation systems. 
Though effective while user-item interactions are abundant, conventional recommendation systems struggle to recommend cold-start items without historical interactions. To address this, we propose utilizing LLMs as data augmenters to bridge the knowledge gap on cold-start items during training. We employ LLMs to infer user preferences for cold-start items based on textual description of user historical behaviors and new item descriptions. 
The augmented training signals are then incorporated into learning the downstream recommendation models through an auxiliary pairwise loss. 
Through experiments on public Amazon datasets, we demonstrate that LLMs can effectively augment the training signals for cold-start items, leading to significant improvements in cold-start item recommendation for various recommendation models.

\end{abstract}


\maketitle

\section{Introduction}

Large language models (LLMs) trained on massive amount of web data, embody comprehensive understandings of the world, and have exhibited remarkable reasoning and generalization abilities
~\cite{anil2023palm,brown2020language}. 
They have revolutionized many application fields, from creative writing, conversational agents, to search engine design~\cite{zhao2023survey,qin2023large,kamalloo2023evaluating}.
With their potential to reason and generalize user preferences and item characteristics better, we explore their application for recommendation systems, especially under cold-start setup. Recommendation systems serve as essential conduits conveying interesting contents to users across a wide range of online platforms. These systems typically operate by analyzing users' historical interactions to infer their preferences and subsequently recommend items that align with those preferences. The most widely adopted model-based recommendation systems heavily rely on well-learned ID-based embeddings \cite{he2017neural,wang2019neural} for both users and items to memorize and predict their relevance. 

While effective when user item interactions are abundant, these ID embedding-based approaches face a critical challenge in recommending fresh and tail content, which lack the initial exposure and interaction data necessary for the model to learn accurate embeddings. This is also known as the \textit{cold-start} problem\footnote{We focus on cold-start item recommendation here and leave recommendations for cold-start users for future work.}. To mitigate this long-standing problem, content-based recommendation systems utilize item meta features to assist item representation learning, i.e., replacing the ID-based item embedding with transformation of meta features, or their combination~\cite{schnabel2022situating,zhao2022improving,gope2017survey}. The hope is to generalize the learning power from items with abundant interactions to cold-start items through shared meta features. 

Recent breakthroughs in LLMs and other foundation models offer exciting prospects for enhancing recommendation systems. Works in \cite{li2023text,li2023gpt4rec,hou2023large,geng2022recommendation} showcase the promise of turning user query and content features into text, and utilizing the generative LLM models to build ID-free recommendation systems. However, it is still necessary to fine-tune large pre-trained modality encoders for recommendation (i.e., for the parameters and architecture) \cite{yuan2023go}, which is a resource-intensive step and requires a significant amount of engineering effort. Additionally, latency in serving LLMs or large foundation models per user request to obtain recommendation results as required by these approaches is often more than the O(100ms) response time expected on recommendation platforms. Therefore it is prohibitively expensive to meet the Query-Per-Second requirement of industrial recommendation systems. 

To transfer the power of LLMs to address the long-standing recommendation cold-start problem, instead of plugging them at the serving phase, we look into their potential in filling in the data gap  during the \textbf{training phase} for current recommendation systems. In previous works \cite{li2023text,li2023gpt4rec,hou2023large}, it is observed that LLMs are capable of understanding users’ behavior (with appropriate prompt) and generating the context for contents of interest to the users (i.e., directly generating the title of the items or related topics). The research questions we are answering is: 1) can we utilize LLMs' reasoning and generalization capabilities to generate synthetic user behaviors on cold-start items (i.e., ask if a user would prefer to watch a new video about ``prompt-engineering'' taught by Andrew Ng or the new TV series acted by Selena Gomez given the user's watching history); 2) does augmenting the training data for cold-start items with these synthetic interactions improve learning for classic recommendation models? With such a model-agnostic approach, we could bypass the slow API calls for LLM inference during serving time and also address the data sparsity issue for cold-start items. Additionally, 
the augmentation procedure as a data pre-processing step can be easily integrated to any industrial-scale recommenation system, providing a scalable approach to exploit LLMs' generalization capabilities. 
Together we make the following contributions in this work: 1) we propose pairwise comparison prompting LLMs to infer a user's preference between an item pair based on the user's historical interactions; 2) we integrate these LLM-generated synthetic user preferences with a pairwise loss as a supplement to the regular recommendation task; 
3) we experiment on two real-world datasets and show that these synthetic user preferences can significantly improve the performance for cold-start items, and LLMs, even with a small model size, can be effectively used as a data augmenter for cold-start recommendation.

\section{Related Work}
\smallskip
\textbf{LLMs for Recommendation Systems}. There is increasing interest in adapting LLMs directly for recommendation \cite{bao2023tallrec,dai2023uncovering,geng2023vip5,hou2023large,li2023gpt4rec,liu2023chatgpt}. Another line of research explores how to use LLMs to improve conventional recommendation systems by generating new features \cite{li2023taggpt,xi2023towards} or improving the encoding of existing features \cite{yu2021tiny}.
The main advantage of augmenting conventional recommendation systems with LLMs through feature engineering or feature encoders lies in its serving efficiency, since directly serving LLMs to retrieve items from millions or even billions of contents on industrial recommendation platform is prohibitively expensive and prevents their wide adoption. Our work aims to augment conventional ID-based recommmendation with synthetic examples generated by LLMs during training, instead of directly serving LLMs. 

\smallskip
\noindent\textbf{Cold-start item recommendation} has been a persistent challenge in the field of recommender systems. Specifically, we focus on the more challenging scenario where cold-start items are newly uploaded and lack any user feedback. Traditionally, recommendation systems have addressed this issue by incorporating side information about items ~\cite{schnabel2022situating} and learning the correlation between well-trained ID-based embeddings and side information \cite{zhao2022improving}.
While recent efforts enable models to learn item embedding with just few interaction \cite{lee2019melu,du2020learn} via meta-learning, these methods still fall short in effectively handling items with no historical interactions. Our work investigates the feasibility of utilizing LLMs to generate synthetic training signals for cold-start items, 
allowing for direct learning of collaborative embeddings from these synthetic signals.

\begin{figure}[t]
\centering
\includegraphics[width=0.46\textwidth]{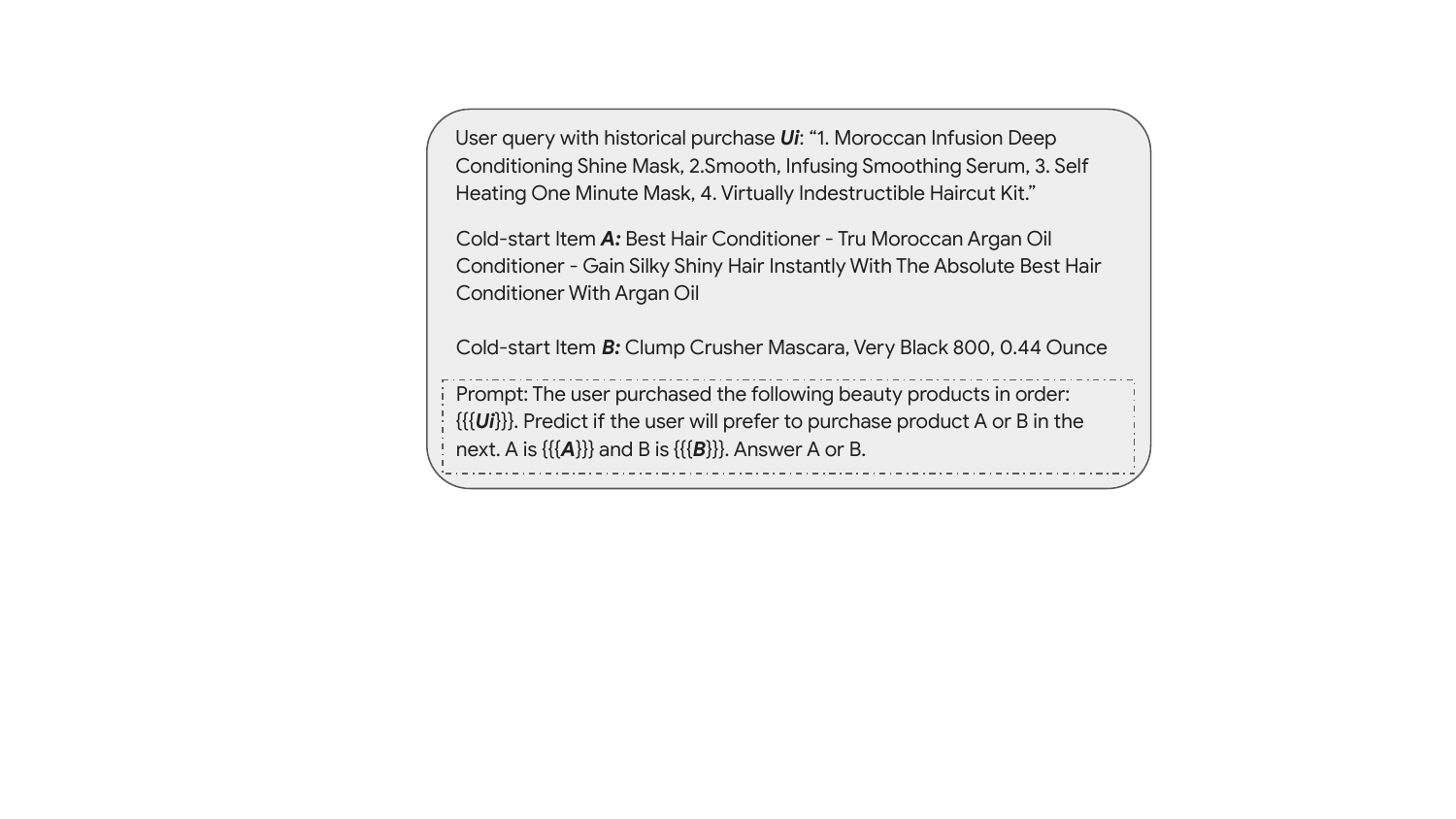}
\vspace{-0.1in}
\caption{Pairwise comparison prompt for a user query.}
\vspace{-0.1in}
\label{fig:prompt}
\end{figure}

\smallskip
\noindent\textbf{Data Augmentation} has demonstrated its effectiveness on enhancing the training efficiency of neural models across various application domains ~\cite{krizhevsky2012imagenet, ding2022data}. In the context of recommendation systems, CLS4Rec, inspired by similar concepts in \cite{devlin2018bert}, employs random item cropping, masking, and reordering to generate augmented views of user historical sequences, ultimately improving model robustness and accuracy. To address the challenge of less active users, \citeauthor{wang2022learning} proposed a learning-to-learn pipeline for augmenting training data and enhancing model performance for this user group. However, our work is the first to generate augmented training data for user behavior understanding based on historical interactions, specifically aiming to bridge the knowledge gap for cold-start items.

\begin{table*}[]
\centering
\caption{Summary of \textbf{cold-start recommendation} and \textbf{warm-start recommendation} performance (\%). The augmented data generated by LLMs significantly boost cold-start recommendation without sacrificing warm-start.} 
\vspace{-0.1in}
\resizebox{\textwidth}{!}{%
\begin{tabular}{cc|cccccc|llllll}
\hline\hline
\multicolumn{2}{c|}{\textit{Task}}                              & \multicolumn{6}{c|}{Cold-start}                                                                                                                & \multicolumn{6}{c}{Warm-start}                                                                                                                                 \\ \hline
\multicolumn{2}{c|}{\textit{Dataset}}                           & \multicolumn{3}{c|}{\textbf{Beauty}}                                             & \multicolumn{3}{c|}{\textbf{Sports}}                      & \multicolumn{3}{c|}{\textbf{Beauty}}                                           & \multicolumn{3}{c}{\textbf{Sports}}                                         \\ \hline
\multicolumn{2}{c|}{\textit{Metrics}}                           & \multicolumn{1}{c|}{R@5} & \multicolumn{1}{c|}{R@10} & \multicolumn{1}{c|}{R@50} & \multicolumn{1}{c|}{R@5} & \multicolumn{1}{c|}{R@10} & R@50 & \multicolumn{1}{c|}{R@5} & \multicolumn{1}{c|}{R@10} & \multicolumn{1}{c|}{R@50} & \multicolumn{1}{c|}{R@5} & \multicolumn{1}{c|}{R@10} & \multicolumn{1}{c}{R@50} \\ \hline


\multicolumn{1}{c|}{\multirow{3}{*}{\textbf{NeuMF}}}  & w/o aug & \multicolumn{1}{c|}{0.14}    & \multicolumn{1}{c|}{0.22}     & \multicolumn{1}{c|}{0.48}     & \multicolumn{1}{c|}{0.01}    & \multicolumn{1}{c|}{0.02}     &  0.13   &      \multicolumn{1}{c|}{3.44}                &     \multicolumn{1}{c|}{5.36}                     & \multicolumn{1}{c|}{18.76}     &             \multicolumn{1}{c|}{2.65}         &            \multicolumn{1}{c|}{4.76}              &       17.98         \\ 
\multicolumn{1}{c|}{}                                 & content-based  & \multicolumn{1}{c|}{0.45}    & \multicolumn{1}{c|}{1.07}     & \multicolumn{1}{c|}{2.13}     & \multicolumn{1}{c|}{0.09}    & \multicolumn{1}{c|}{0.19}     &   0.87   &      \multicolumn{1}{c|}{2.48}            &     \multicolumn{1}{c|}{4.02}                     & \multicolumn{1}{c|}{16.89}     &           \multicolumn{1}{c|}{1.77}             &            \multicolumn{1}{c|}{3.23}              &       16.01                   \\
\multicolumn{1}{c|}{}                                 & w/ aug  & \multicolumn{1}{c|}{1.19}    & \multicolumn{1}{c|}{2.32}     & \multicolumn{1}{c|}{10.53}     & \multicolumn{1}{c|}{0.22}    & \multicolumn{1}{c|}{0.41}     &   2.11   &      \multicolumn{1}{c|}{3.35}            &     \multicolumn{1}{c|}{5.21}                     & \multicolumn{1}{c|}{18.03}     &           \multicolumn{1}{c|}{2.32}             &            \multicolumn{1}{c|}{4.75}              &       17.67                   \\ \hline
\multicolumn{1}{c|}{\multirow{3}{*}{\textbf{SASRec}}} & w/o aug & \multicolumn{1}{c|}{0.18}    & \multicolumn{1}{c|}{0.48}     & \multicolumn{1}{c|}{0.74}     & \multicolumn{1}{c|}{0.10}    & \multicolumn{1}{c|}{0.22}     &   0.31   &          \multicolumn{1}{c|}{4.25}                &               \multicolumn{1}{c|}{6.18}           & \multicolumn{1}{c|}{19.87}     &          \multicolumn{1}{c|}{3.57}               &        \multicolumn{1}{c|}{5.48}                  &          19.01                \\
\multicolumn{1}{c|}{}                                 & content-based  & \multicolumn{1}{c|}{0.56}    & \multicolumn{1}{c|}{1.26}     & \multicolumn{1}{c|}{4.77}     & \multicolumn{1}{c|}{0.15}    & \multicolumn{1}{c|}{0.30}     &   0.96   &      \multicolumn{1}{c|}{2.90}            &     \multicolumn{1}{c|}{5.01}                     & \multicolumn{1}{c|}{16.97}     &           \multicolumn{1}{c|}{1.89}             &            \multicolumn{1}{c|}{3.09}              &       16.33                            \\

\multicolumn{1}{c|}{}                                 & w/ aug  & \multicolumn{1}{c|}{1.34}    & \multicolumn{1}{c|}{2.47}     & \multicolumn{1}{c|}{11.40}     & \multicolumn{1}{c|}{0.37}    & \multicolumn{1}{c|}{0.61}     &    2.41   &         \multicolumn{1}{c|}{4.30}                 &        \multicolumn{1}{c|}{6.11}                   & \multicolumn{1}{c|}{19.79}     &         \multicolumn{1}{c|}{3.51}                 &            \multicolumn{1}{c|}{5.39}               &         18.95               \\ \hline\hline

\end{tabular}%
}
\label{tab:results}
\end{table*}

\section{Preliminaries}
Let $\mathcal{U}=\{u_1, u_2, ..., u_G\}$ represent the user set, $\mathcal{I}_{warm}=\{i_1, i_2, ..., i_P\}$ and  $\mathcal{I}_{cold}=\{i_{P+1}, i_{P+2}, ..., i_{P+N}\}$ represent the warm and cold-start items on the platform respectively.
Each item is mapped to a trainable embedding associated with its ID. The principle of generating personalized recommendation is to predict the compatibility between users and items, from which the items with high compatibility to a user would be retrieved from a large set of candidate items to make up the unique recommendation list for the user. As one of the most effective variants of matrix factorization \cite{he2017neural}, latent factor model-based recommendation has attracted lots of attention due to the Netflix Prize. The high level idea of latent factor models is to approximate compatibility between a user and an item with the dot product of the corresponding latent factor vectors. Given that $\mathbf v_{u}$ and $\mathbf v_{i}$ denote the latent factor vector for user $u$ on item $i$, a latent factor model calculates their compatibility via $\hat{y}_{u,i} = \mathbf v_{u}^{T}\mathbf v_{i}$.
Most of the widely-adopted recommendation frameworks can be regarded as extension of such latent factor model. Cold-start items, of course, have no training signal to obtain informative embeddings $\mathbf v_{i}$. To bridge the knowledge gap, we propose to generate synthetic data to simulate users' interactions on the cold-start items.

\section{LLMs as Data Augmenters}
\smallskip
\textbf{Augmented Data Generation.} We focus on the PaLM family \cite{anil2023palm}, and directly use their generation without any finetuning. We follow \cite{hou2023large} and simply put the description of items that the user has interacted with into the prompt. Specifically, given a user query $U_{i}$ in the training set, we adopt the descriptive item titles to denote each historical interactions. 
To infer users' preferences from this descriptive user query, we can either ask the user if he/she would like a specific cold-start item (pointwise) or ask them if he/she prefer cold-start item A or B (pairwise). LLMs have been shown to struggle with calibrated pointwise relevance estimation \cite{ouyang2022training}, but demonstrate better capabilities \cite{dai2023uncovering,qin2023large} at pairwise comparison task. Therefore, we probe LLMs to generate pairwise preference between cold-start items given a user query. In particular, we randomly sample an item pair $(A, B)$ with $A,B\in\mathcal{I}_{cold}$, and construct the prompt as shown in Figure \ref{fig:prompt} to retrieve the user's preference between $A$ and $B$. 
The pairwise comparison in the prompt ensures that we obtain training signal from every LLM call to indicate preference between two cold-start items for the user. In comparison, a pointwise inference on a random cold-start item for the user will most likely result in negative labels (not interested) while positive labels are rare.

\smallskip
\noindent\textbf{Pairwise Comparison Loss.} To incorporate this augmented signal during the training process, we add the pair-wise preference prediction on the cold-start item pairs as an auxiliary task complementary to the regular recommendation task. The answer returned by the LLM would be regarded as the $pos$ item and the other item as the $neg$. And given that a user $u$ prefers cold-start item $pos$ than cold-start item $neg$, we have the following pairwise loss inspired by Bayesian Personalized Ranking (BPR) loss  \cite{rendle2012bpr}:
\begin{equation}
    \mathcal{L}_{aug} = -\sum_{(u, pos, neg)}\ln\sigma(\hat{y}_{u,pos} - \hat{y}_{u,neg}),
\label{equ:loss}
\end{equation}
in which $\sigma$ is the Sigmoid activation function. 
We then add this pairwise BPR loss to the sampled batch softmax \cite{yi2019sampling} loss typically used in training recommendation model.
The pairwise loss back-propagates gradients to the embeddings of the positive and negative cold-start items, which are often under-trained due to the lack of interaction, and thus training signals from the main task.

\section{Experiments}


\subsection{Experimental Setup}
\noindent\textbf{Data and Preprocessing.}
We use the public Amazon review datasets \cite{ni2019justifying} for evaluating the performance of our method. We select the ``\textit{Beauty}'' and ``\textit{Sports and Outdoors}'' categories, which include users' ratings and reviews on items falling into these two categories. To split the datasets for training and testing, we follow the single-time-point split \cite{sun2023take} and select the time-point to split the data by $7$:$3$. The interaction data before the splitting time-point would be used for model training and the trained model is tested on interaction data after the splitting time-point. Specifically, we regard the items \textit{showing up only in the testing data} as \textit{cold-start items} and others as \textit{warm-start} items, which is similar to the real-world setup. As an example, the split resulted in 55,255 warm-start items and 2,751 cold-start items in the ``\textit{Sports and Outdoors}'' category. There are 224,956 user queries in this category. We randomly sample a subset of user queries, and randomly sampled two cold-start items for each query to generate augmented data examples.

\smallskip
\noindent\textbf{Models and Parameters.}
We evaluate the generalizability of our proposed methods  employing two established recommender backbones: Neural Matrix Factorization (\textbf{NeuMF}) \cite{he2017neural} and \textbf{SASRec} \cite{kang2018self}. NeuMF learns user embedding $\mathbf v_{u}$ and item embedding $\mathbf v_{i}$ from their IDs, and SASRec leverages a self-attention layer over user's historical interactions to encode sequential information in user embedding $\mathbf v_{u}$. These backbones represent core components in many recommender systems, and we test three variations built upon them: 1) Baseline (\textbf{\textit{w/o Aug}}) trains recsys with the original training data; 2) \textbf{\textit{content-based}} method employs bag-of-words~\cite{zhang2010understanding} representations for items, incorporating their category and title, which is a common approach for handling cold-start items and 3) LLM-augmentation (\textbf{\textit{w Aug}}) incorporates LLM-generated augmentations and supplement the training process with pairwise comparison loss in Equ \ref{equ:loss}. We use \textbf{\textit{PaLM2}} \cite{anil2023palm} with different model size (i.e., \textit{XXS}, \textit{S} and \textit{L}) to investigate the performance on the synthetic data generation.


\smallskip
\noindent\textbf{Top-K Evaluation Metrics.} For each test user query, we retrieve the top-K items from $\mathcal{I} = \mathcal{I}_{cold}\cup \mathcal{I}_{warm} $  with the highest compatibility scores. And to compare the performance of the recommendation systems offline, we adopt recall (i.e., \textit{R@K}) to check if the ground-truth item of the testing query shows up in the top-K list. 
We group the results by ground-truth label, i.e., grouping results when the purchased items are cold-start items in the "cold-start" columns, and the rest in the "warm-start" columns in Table~\ref{tab:results}. 

\subsection{Results and Analysis}
In the overall comparison, we use \textit{PaLM2-S} to generate the synthetic data and combine it with the regular training data. We randomly sample 20\% of user queries and generate one augmented data example for each query.
In Table \ref{tab:results}, we report the results to test its effectiveness on both NeuMF and SASRec. Without any augmentation data examples, cold-start items lack training signals, leading to extremely low recall at all K values. While content-based leveraging item titles and descriptions can offer decent performance on cold-start items, it neglects collaborative signals, significantly hurting warm-start item recommendations. In contrast, the augmented training signals for cold-start items, when learned through the pairwise loss, can benefit the representation learning for both NeuMF and SASRec, and boost the performance on cold-start recommendation significantly. Furthermore, the augmented signals and pairwise comparison loss improve recall more at higher K values, as they enable the model to rank a wider range of cold-start items, including some less relevant ones (i.e., which still match users' preferences).  
The results suggest that LLMs is an effective data augmenter to fill up the missing knowledge on cold-start items. Although the augmented training signals for cold-start items hurt the warm-start item recommendations slightly for some recall metrics, the drop in performance is marginal compared to the huge gain obtained on cold-start recommendations. 

\begin{figure}
    \centering
    \subfigure[]
    {
    \includegraphics[width=0.48\columnwidth]{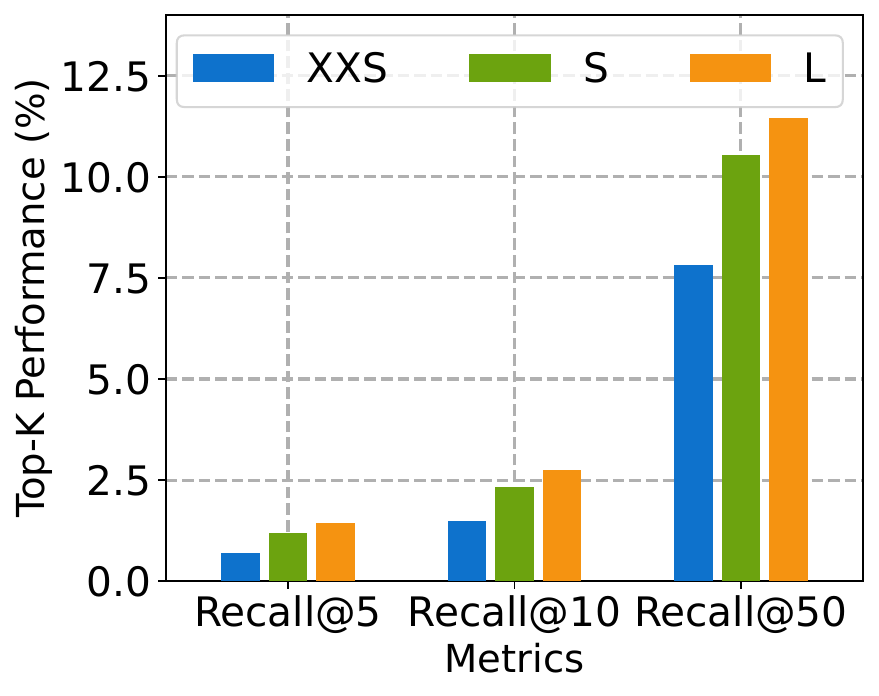}
    }
    \hspace{-0.15in}
    \subfigure[]
    {
    \includegraphics[width=0.48\columnwidth]{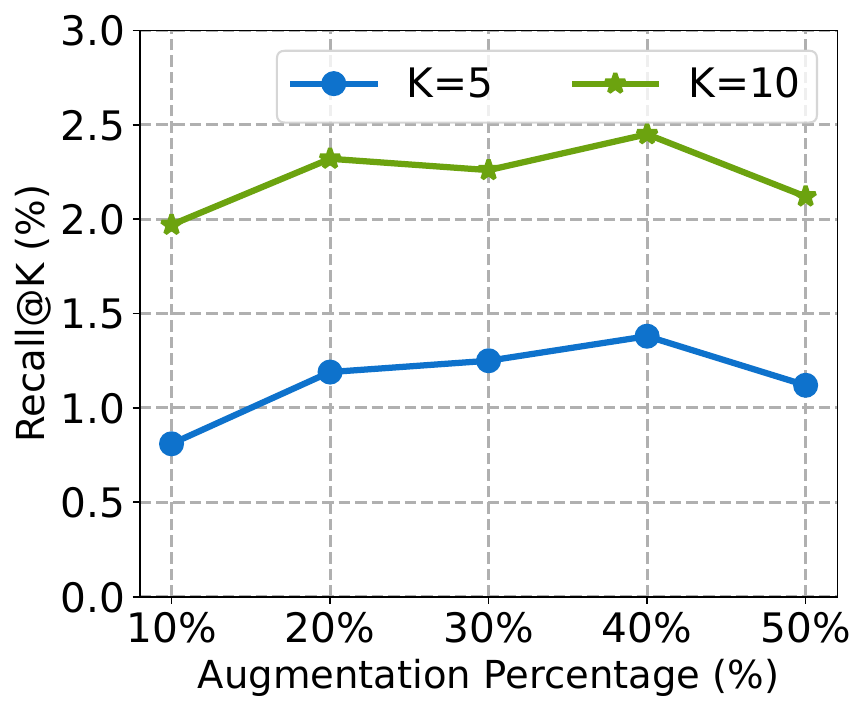}
    }
    \vspace{-0.15in}
    \caption{Cold-start Recommendation Under Different Model Sizes and Augmentation Percentage.}
    \label{fig:size}
    \vspace{-0.1in}
\end{figure}

To understand the impact of different LLM model sizes, we conduct experiments on Amazon-Beauty using NeuMF based recommender with \textit{PaLM2} of various model sizes (i.e., XS, S and L) as the data augmenter. In Figure \ref{fig:size} (a), we find that the model size does influence the performance of augmentation. It is known that many abilities of LLMs are emergent as these models scale up \cite{anil2023palm,brown2020language}. 
We hypothesize larger models are able to reason through the user historical behaviors better and infer the preferences more accurately. 
In Figure \ref{fig:size} (b), we also observe that by generating augmented training signals with more user queries, we can further increase the performance for cold-start recommendation. Even though adding more synthetic data beyond certain point ($40\%$) did not lead to any further improvement. 



\section{Conclusion}
Addressing the lack of user interactions with cold-start items is crucial for enhancing recommendation effectiveness. We propose to employ LLMs to generate augmented training signals for cold-start items in recommender systems. We use a pairwise comparison prompt to leverage LLMs to infer user preferences between pairs of cold-start items. This model-agnostic design provides informative training signals for cold-start items without introducing additional computational overhead during serving time. Experiments on public datasets show that our method generates effective augmented training signals and improves cold-start item recommendation.

\bibliographystyle{ACM-Reference-Format}
\bibliography{sample_bib}

\end{document}